\input{aipcheck}

\documentclass[
    ,final            
  ]
  {aipproc}

\layoutstyle{6x9}

\def\simlt{\mathrel{\rlap{\lower 3pt\hbox{$\sim$}}\raise 2.0pt\hbox{$<$}}}
\def\simgt{\mathrel{\rlap{\lower 3pt\hbox{$\sim$}} \raise 2.0pt\hbox{$>$}}}
\def\lta{\mathrel{\rlap{\lower 3pt\hbox{$\sim$}}\raise 2.0pt\hbox{$<$}}}
\def\gta{\mathrel{\rlap{\lower 3pt\hbox{$\sim$}} \raise 2.0pt\hbox{$>$}}}

\begin{document}

\title{Supermassive black hole mergers and cosmological structure formation}


\author{Marta Volonteri}{
  address={Institute of Astronomy, Madingley Road, Cambridge, UK}
}

\begin{abstract}
Massive black holes (MBHs), with masses in the range $10^3-10^8\,M_\odot$, which merge with a companion black hole of similar mass are expected to be the most powerful source of gravitational radiation in the frequency range probed by LISA. MBH binaries can be detected by LISA up to $z\simeq 5-15$. Gravitational waves from MBH mergers can serve as a powerful tool to study the early evolution of the MBH population, and possibly the role played by MBHs in the joint evolution with their hosts. I review scenarios for the co-evolution of MBHs and cosmological structures, where MBH seeds form in pre-galactic structures. These black holes evolve then in a hierarchical fashion, following the merger hierarchy of their hosts. Accretion of gas, traced by quasar activity, plays a fundamental role in determining the two parameters defining a black hole, mass and spin. Gravitational waves, together with observations in electromagnetic bands, can help constrain the evolution of both MBH mass and spin.
\end{abstract}

\maketitle

\section{Introduction}
Massive black holes, with masses above thousands solar masses,  have existed in the Universe since early times.  We can easily trace their presence, as massive black holes are the engines powering the luminous quasars that have been detected up to high redshift. Nowadays we can detect in neighboring galaxies the dead remnants of this bright past activity. It is well established observationally that the centers of most galaxies host MBHs with masses in the
range $M_{BH} \sim 10^6-10^9\,M_\odot$ (e.g., Ferrarese \& Merritt 2000; Kormendy \& Gebhardt 2001; Richstone 2004). The MBH population may extend down to the smallest masses.  For example, the dwarf Seyfert~1 galaxy POX 52 is thought to contain a BH of mass $M_{BH} \sim 10^5\,M_\odot$ (Barth et al. 2004). At the other end, however, the Sloan Digital Sky survey detected luminous quasars at very high redshift, $z>6$. Follow-up observations confirmed that at least some of these quasars are powered by supermassive black holes with masses $\simeq 10^9\, M_\odot$ (Barth et al. 2003; Willott et al. 2003). 
The mass of the quiescent MBHs detected in the local Universe scales with the bulge luminosity - or stellar velocity dispersion - of their host galaxy (Ferrarese \& Merritt 2000; Gebhardt et al. 2000; Tremaine et al. 2002), suggesting a single mechanism for assembling black holes and forming spheroids in galaxy halos. The evidence is therefore in favor of a co-evolution between galaxies, black holes and quasars.  

In cold dark matter cosmogonies, small-mass subgalactic systems form first to merge later into larger and larger structures. In this paradigm galaxy halos experience multiple
mergers during their lifetime. If every galaxy with a bulge hosts a MBH in its center, 
and a local galaxy has been made up by multiple mergers, then a black hole binary is a natural 
evolutionary stage. The investigation of the dynamical and gravitational processes involving MBH binaries, as well as their occurrence and distribution  clearly has to take into account the cosmological framework and the dynamical evolution of both MBHs and their hosts.  

\section{Scenarios for massive black hole formation}
The formation of massive black holes is far less understood than that of their light, stellar mass, counterparts. The "flowing chart" presented Rees (1978) still stands as a guideline for the possible paths leading to formation of massive BH seeds in the center of galactic structures. One first possibility is the direct formation of a BH from a collapsing gas cloud (Haehnelt \& Rees 1993; 
Loeb \& Rasio 1994; Eisenstein \& Loeb 1995; Bromm \& Loeb 2003; Koushiappas, Bullock \& Dekel 2004; Begelman, Volonteri \& Rees 2006; Lodato \& Natarajan 2006). The main issue for this family of models is how to shed the angular momentum of the gas, so that the gas can collect in the very inner region of the galaxy. The loss of angular momentum can be driven either by (turbulent) viscosity or by global dynamical instabilities, such as the "bars-within-bars" mechanism (Shlosman, Frank \& Begelman 1989).  The gas can therefore condense to form a central massive object, either a supermassive star, which eventually becomes subject to post-Newtonian gravitational instability and forms a seed BH, or via a low-entropy star-like configuration where a small black hole forms in the core and grows by accreting the surrounding envelope.  The mass of the seeds predicted by different models vary, but typically are in the range $M_{BH} \sim 10^4-10^6\,M_\odot$. 

Alternatively, the seeds of MBHs can be associated with the remnants of the first generation of stars, formed out of zero metallicity gas. The first stars are believed to form at $z\sim 20$ in halos which represent high-$\sigma$ peaks of the primordial density field.  The main coolant, in absence of metals, is molecular hydrogen, which is a rather inefficient coolant.  The inefficient cooling might lead to a very top-heavy initial stellar mass function, and in particular to the production of very massive stars with masses $>100 M_\odot$ (Carr, Bond, \& Arnett 1984). If very massive stars form above 260 $M_\odot$,  they would rapidly collapse to massive BHs
with little mass loss (Fryer, Woosley, \& Heger 2001), i.e., leaving behind seed BHs with masses $M_{BH} \sim 10^2-10^3\,M_\odot$ (Madau \& Rees 2001).  

Current observations are unable to distinguish among these (or other) scenarios. The initial conditions are mostly washed out when MBHs gain most of their mass, between $z=3$ and $z=1$. Detection of gravitational waves from merging MBHs at early times can be of paramount importance to understand the preferred path for MBH seed formation (see section 6). 

\section{Dynamical evolution  of massive black holes: gas rich and gas poor hosts}
In the hierarchical structure formation scenario galaxies forms through a series of mergers, which involve the dark matter halo and the internal structure, that is gas, stars, and the central MBH, if any. 
The merging timescale for the galaxies is determined mainly by dynamical friction against the dark matter, which comprises most of the mass of the systems. When two comparable-mass halos (``major mergers'') merge, dynamical friction drags the satellite, along with its central MBH, towards the center of the more massive progenitor.  The process eventually leads to the formation of a bound MBH binary with separation of $\sim$ pc (Yu 2002, Mayer et al 2006, Dotti, Colpi \& Haardt 2006).  One important piece of information which is still missing, is the typical timescale for a BH binary to merge. After dynamical friction ceases to be efficient, at parsec scale or so, there is still 
a large gap before emission of gravitational waves can be efficient in bringing the binary to coalescence 
in less than a Hubble time. 

In massive galaxies at low redshift, the subsequent evolution of the 
binary may be largely determined by three-body interactions with background stars
(Begelman, Blandford \& Rees 1980; see Merritt \& Milosavljevic 2005 for a review). Dark matter particles will be ejected by decaying binaries in the same way as the stars. 
Another possibility is that gas processes, rather than three-body interactions with stars or DM, may induce MBH binaries to shrink rapidly and coalesce (e.g. Mayer et al. 2006; Dotti et al. 2006; Escala et al. 2004; Armitage \& Natarajan 2005; Gould \& Rix 2000). If stellar dynamical or gaseous processes drive the binary sufficiently close ($\lta 0.01$ pc), gravitational radiation will eventually dominate angular momentum and energy losses and cause the
two MBHs to coalesce. 

In gas rich high redshift halos, the orbital evolution of the central MBH binary is likely dominated by dynamical friction against the surrounding gaseous medium. The available simulations  (Mayer et al. 2006, Dotti et al. 2006, Escala et al. 2004) show that the binary can shrink to about parsec or slightly subparsec scale by dynamical friction against the gas, depending on the gas thermodynamics. These binary separations are still too large for the binary to coalesce within the Hubble time owing to emission of gravitational waves. On the other hand, the interaction between a binary and an accretion disc can lead to a very efficient transport of angular momentum. The secondary MBH can therefore reach the very subparsec separations at which emission of gravitational radiation dominates on short timescales (Armitage \& Narajan 2005; Gould \& Rix 2000).  

\begin{figure}
\centering
\includegraphics[height=.5\textheight]{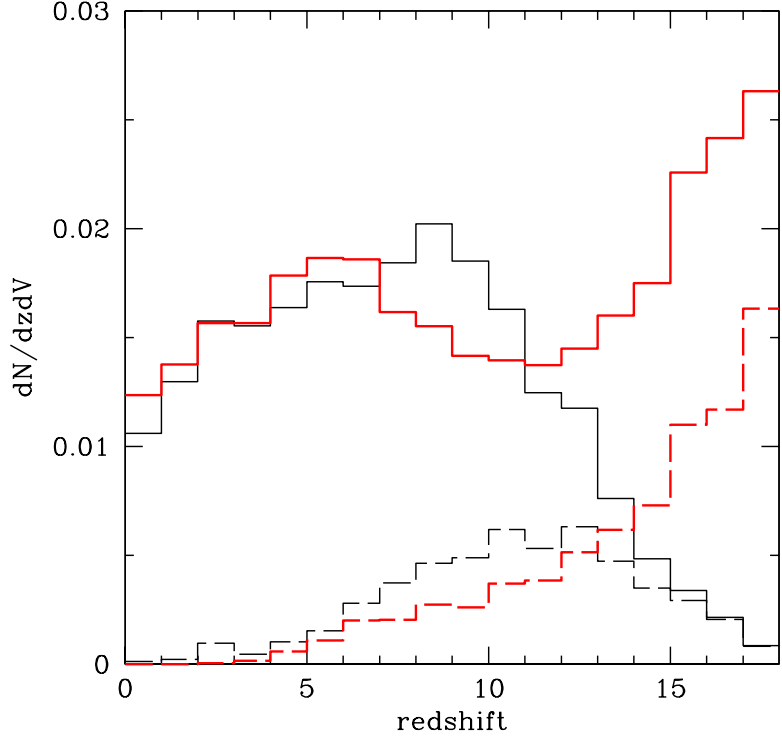}
\caption{Rates of binary MBH mergers ({\it solid line}) and ejections due to the 
gravitational radiation recoil ({\it dashed line}) per comoving Mpc$^3$. The 
subset of ejected binaries is selected requiring that the recoil velocity is larger than 
the escape velocity from the host halo. At redshifts $z>10$ about 80\% of the merging MBH 
binaries are ejected, and therefore lost, into the intergalactic medium. The thin black lines refer 
to a model with inefficient BH mergers, as binaries shrink by stellar dynamics. The thick red lines 
show the case in which BH binaries evolve on the viscous timescale, if an accretion disc is present (\emph{efficient merging}).}
\label{mvfig:4}       
\end{figure}

Figure \ref{mvfig:4} compares the MBH merger rates in two different models accounting for the orbital evolution of MBH binaries in the phase preceding emission of gravitational waves.  We have compared  a conservative \emph{inefficient merging} case in which the interaction with gas is neglected, to a \emph{efficient merging} model (Volonteri \& Rees 2006).  The latter assumes  that, if an accretion disc is surrounding a hard MBH binary, the MBHs coalesce instantaneously owing to interaction with the gas disc. If instead there is no gas readily available, the binary will be losing orbital energy to the stars and the dark matter background.  

\section{Slingshots and rockets: massive black hole ejections}
The lifetime of BH binaries can be long enough (Begelman, Blandford \& Rees 1980; 
Quinlan \& Hernquist 1997; Milosavljevic \& Merritt 2001; 
Yu 2002) that following another galactic merger a third BH can fall in and disturb the 
evolution of the central system.
The three BHs are likely to undergo a complicated resonance scattering interaction, leading to 
the final expulsion of one of the three bodies (`gravitational slingshot') and to the recoil of 
the binary. Any slingshot, in addition, modifies the binding energy of the binary, typically
creating more tightly bound systems.

Another interesting gravitational interaction between black holes happens
during the last stage of coalescence, when the leading physical 
process for the binary evolution becomes the emission of gravitational waves. 
If the system is not symmetric (e.g. BHs have unequal masses or spins) there would be a recoil due to 
the non-zero net linear momentum carried away by gravitational waves in the coalescence 
(`gravitational rocket'). The recoil velocity during the plunge phase probably has the largest contribute. Calculations of the gravitational recoil inside the innermost stable circular orbit (ISCO) naturally have large uncertainties.
Blanchet et al. 2005 calculated the gravitational recoil at the second post-Newtonian 
order for non-spinning holes.  Damour \& Gopakumar 2006, using the effective one-body approach, predict velocities about a factor of 3 less than  Blanchet et al. The latest estimate on the recoil comes from fully relativistic numerical simulations (Baker et al. 2006) following
the dynamical evolution of a black hole binary within the ISCO. These simulations, carried for a mass ratio $q=0.2$, predict a recoil midway through the Blanchet et al. 2005 and Damour \& Gopakumar 2006, predictions. The recoil predicted by Baker et al. 2006 is still large enough to eject the merging binary from small pre-galactic structures.  

Slingshots and rockets basically give BHs a recoil velocity that can even exceed the escape 
velocity from the host halo and spread BHs outside galactic nuclei (Volonteri, Haardt \& Madau 2003). 
In the shallow potential wells of mini-halos, the growth of BHs from seeds can be halted by these
ejections (Haiman 2004; Yoo \& Miralda-Escud\'e 2004). Volonteri \& Rees (2006) estimate that up to $50-80\%$ of black holes merging in high-redshift halos can be ejected due to the gravitational rocket effect. 

\section{Massive black hole spins}
\begin{figure}
\centering
\includegraphics[height=.5\textheight]{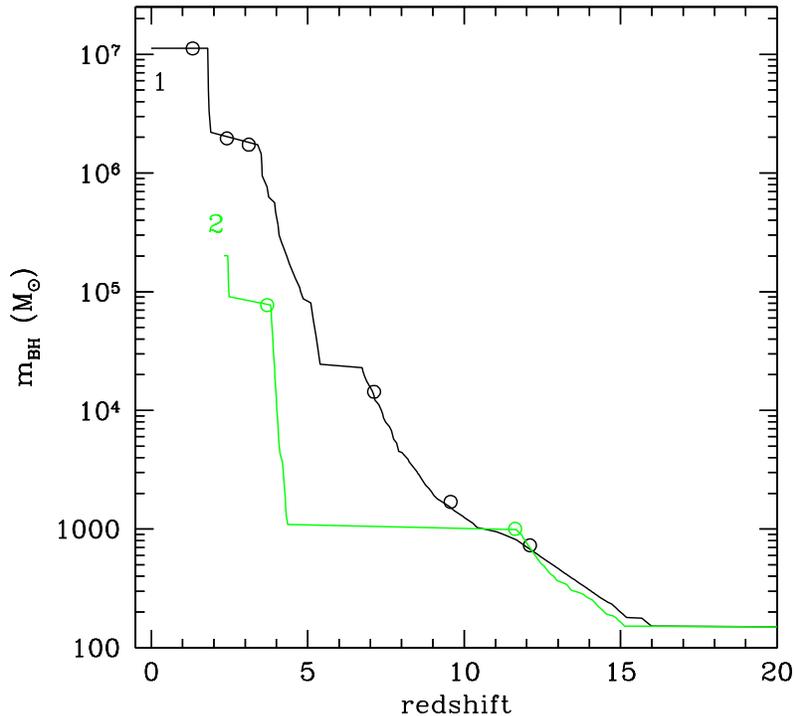}
\caption{Mass-growth history of two MBHs, one ending in a massive 
halo (`1') at $z=0$, and one in a satellite (`2') at $z=2.3$. 
The MBH mass grows by gas accretion after major merger events and by MBHs mergers ({\it circles}).  
Note how most of the mass of the black holes is gained in accretion episodes and not through MBH mergers.}
\label{mvfig:2}       
\end{figure}

The spin of a MBH is modified by a merger with a secondary MBH, or by interplay with the accretion disc and magnetic fields. If accretion proceeds from a thin disk, and magnetic processes are not important, rapidly spinning MBHs are to be expected (Volonteri et al. 2005, Moderski et al. 1998). BH mergers, in fact, do not lead to a systematic effect, but simply cause the  BHs to random-walk around the spin parameter they had at birth.  BH spins, instead, efficiently couple with the angular momentum of the accretion disk, producing Kerr holes independent of the initial spin. This is because, for a thin accretion disk, the BH aligns  with the outer disk on a timescale that is much shorter than the Salpeter time (Natarajan \& Pringle 1998), corresponding to an e-folding time for accretion at the Eddington rate, leading to most of the accretion being from prograde equatorial orbits.  

We can expect that if MBHs have grown in mass mainly through mergers of small seeds, the spin distribution is random, or peaked at the typical spin parameter of MBH seeds, if any. The comparison between the total mass density accreted by MBHs, as traced by the luminosity function of quasars, and the  mass density that we see in MBHs today (Soltan's argument), though, suggests that the mass build-up is dominated by accretion  (Yu \& Tremaine 2002; Elvis, Risaliti, \& Zamorani 2002; Marconi et  al. 2004), with mergers playing a secondary role (Figure 2). Several theoretical models have proved equally successful in explaining the evolution of MBHs and quasars in the redshift range $1\lta z\lta 3$ (e.g., Kauffmann \& Haehnelt 2000; Wyithe \& Loeb 2002, 2003; Cattaneo et al. 1999). In this theoretical models, during the quasar peak epoch most accretion episodes typically double the MBH mass.  If all this matter is accreted by the MBHs at a high rate, via a radiatively efficient thin disc, then most of the mass accreted by the hole acts to spin it up, even if the orientation of the spin axis changes in time. Most individual accretion episodes thus produce rapidly-spinning MBHs independent of the initial spin.

\begin{figure}
\centering
\includegraphics[height=.5\textheight]{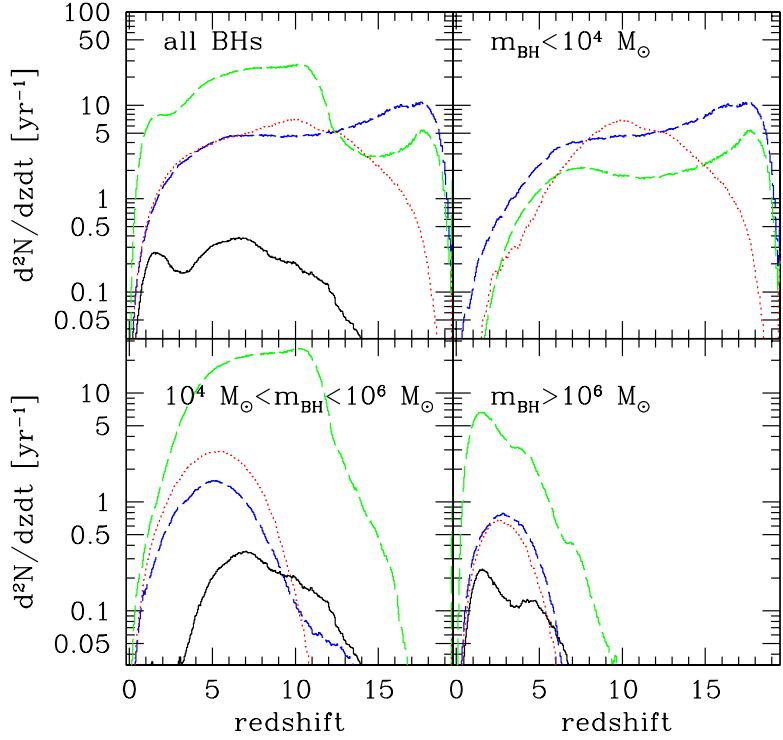}
\caption{Predictions on the number of MBH binary coalescences observed by LISA per year at $z=0$, 
per unit redshift, in different BH mass intervals. If MBHs form from very massive seeds at low redshift (see Haehnelt 2004), the expected merger rate is of order a few per year (cfr. bottom panel). If MBHs evolve from high redshift small seeds the merger rate can be up to tens or hundred per year (cfr. upper panels). {\it Dotted curve:} MBH seeds from PopIII stars, hardening in a stellar background (see Sesana et al. 2004, 2005). {\it Solid curve:} MBH seeds from PopIII stars, when an accretion disc is present, the satellite MBH is dragged towards the central MBH on the viscous timescale (\emph{efficient merging}). {\it Long dashed curve:} MBH seeds from direct collapse (Koushiappas et al. 2004), \emph{efficient merging}.  {\it Short dashed curve:} MBH seeds from direct collapse (Begelman, Volonteri \& Rees 2006, Sesana et al. 2006), \emph{efficient merging}.}
\label{mvfig:1}      
\end{figure}

\section{Conclusions}
The evolution of the MBH population is long and complicated. Electromagnetic observations have provided constraints on the accretion properties of MBHs, and on the evolution of the population as a whole, mainly at $z\lta 3$. Alhough we start having a coherent picture on how the population of MBHs evolved, there are several crucial issues that remain to be clarified. I'll focus here on two points, namely the formation of MBH seeds and the evolution of MBH spins, where the contribution of gravitational radiation can be essential. 

Measuring MBH spin parameters is a very arduous task, and all evidences are for the moment indirect. 
A promising technique is based on the shape of the iron line which can be detected in the X-ray spectrum of accreting MBHs, and has probably been observed in some local Seyfert galaxies (Miniutti, Fabian \& Miller 2004, Streblyanska et al. 2005). The location of the inner radius of the accretion disc (corresponding to the ISCO in the standard picture) has a large impact on the shape of the line profile,  with small inner disc radii implying stronger relativistic effects which modify the expected line profile.  The detection and modelling of a relativistic iron line via X-ray observations can therefore potentially provide a way to measure MBH spins.  The assumption that the inner disc radius corresponds to the ISCO is not a trivial one, however, especially for thick discs (e.g. Krolik 1999, but see Afshordi \& Paczynski 2003), and the detection and identification of the iron line is uncertain in most sources.  The detection of gravitational waves from merging binaries can provide a unique tool to constrain MBH spins.  Indeed the spin is a measurable parameter, with a very high accuracy, in the gravitational waves LISA signal (Barack \& Cutler 2004, Berti et al. 2004, Lang \& Hughes 2006, Vecchio 2004, Berti et al. 2006) 

When the early evolution of MBHs is concerned,  current observations are still far away from the possibility of observing miniquasars powered by MBHs with mass of a few thousands solar masses at $z>10$.  Some constraints on the global accretion history, even at high redshift,  can be already put by comparing theoretical models predictions to ultra-deep X-ray surveys (Salvaterra et al. 2006). Future X--ray missions,  such as {\it XEUS} and {\it Constellation-X}, and near infrared  facilities such as {\it JWST}, will have the technical capabilities to detect accreting MBHs at $z\simgt 6$ down to a mass limit as low as $10^5-10^6 M_\odot$, giving constraints on the accretion properties of MBHs at early times. The masses of the putative MBH seeds however are in the range $10^2-10^6\; M_\odot$, so still beyond the capabilities of planned X--ray  and near infrared missions. 

LISA in principle is sensible to gravitational waves from binary MBHs with masses in the range $10^3-10^6\; M_\odot$ out to $z\simeq 10$. Different theoretical models for the formation of MBH seeds and dynamical evolution of the binaries predict merger rates that largely vary one from the other. Figure 3 shows a comparison of the theoretical merger rates predicted when considering either different seed models or variations in the dynamical evolution of MBH binaries (see sections 2 and 3 above, and Sesana et al. 2006).

\begin{theacknowledgments}
I wish to acknowledge here several of my collaborators, Mitch Begelman, Francesco Haardt, Piero Madau, Martin Rees, and Alberto Sesana. 
\end{theacknowledgments}

\end{document}